\documentclass[%
reprint,
superscriptaddress,
 amsmath,amssymb,
 aps
]{revtex4-2}

\usepackage{graphicx}%
\usepackage{multirow}%
\usepackage{amsmath,amssymb,amsfonts}%
\usepackage{amsthm}%
\usepackage{mathrsfs}%
\usepackage[title]{appendix}%
\usepackage{xcolor}%
\usepackage{textcomp}%
\usepackage{booktabs}%
\usepackage{listings}%
\usepackage{indentfirst}
\usepackage{float}
\usepackage{hyperref}
\usepackage[all]{hypcap}
\hypersetup{
    colorlinks=true,
    linkcolor=blue,
    filecolor=blue,      
    urlcolor=blue,
    citecolor=blue
    }

\usepackage{lineno} 
\usepackage{soul} 

\newcommand{\ket}[1]{\left|{#1}\right\rangle}

\newcommand{\brak}[1]{\left\langle{#1}\right\rangle}

\let\up\uparrow
\let\down\downarrow

\newcommand{\beq}{\begin{equation}}
\newcommand{\eeq}{\end{equation}}
\newcommand{\beqa}{\begin{eqnarray}}
\newcommand{\eeqa}{\end{eqnarray}}

\usepackage{titlesec}
\titleformat*{\section}{\centering\footnotesize\bfseries\uppercase}

\usepackage[paperwidth=210mm,paperheight=297mm,centering,hmargin=1.7cm,vmargin=1.5cm]{geometry}

\begin{document}

\nolinenumbers

\title{\large \textbf{Observing Spatial Charge and Spin Correlations\\in a Strongly-Interacting Fermi Gas}}

\author{Cyprien Daix}
\affiliation{Laboratoire Kastler Brossel, ENS-Universit\'{e} PSL, CNRS, Sorbonne Universit\'{e}, Coll\`{e}ge de France, 24 rue Lhomond, 75005, Paris, France}
\author{Maxime Dixmerias}
\affiliation{Laboratoire Kastler Brossel, ENS-Universit\'{e} PSL, CNRS, Sorbonne Universit\'{e}, Coll\`{e}ge de France, 24 rue Lhomond, 75005, Paris, France}
\author{Yuan-Yao He}
\thanks{Correspondence to be addressed to: \href{mailto:tarik.yefsah@lkb.ens.fr}{tarik.yefsah@lkb.ens.fr}, \href{mailto:szhang@flatironinstitute.org}{szhang@flatironinstitute.org}, and \href{mailto:heyuanyao@nwu.edu.cn}{heyuanyao@nwu.edu.cn}}
\affiliation{Institute of Modern Physics, Northwest University, Xi'an 710127, China}
\affiliation{Shaanxi Key Laboratory for Theoretical Physics Frontiers, Xi'an 710127, China}
\affiliation{Peng Huanwu Center for Fundamental Theory, Xi'an 710127, China}
\author{Joris Verstraten}
\affiliation{Laboratoire Kastler Brossel, ENS-Universit\'{e} PSL, CNRS, Sorbonne Universit\'{e}, Coll\`{e}ge de France, 24 rue Lhomond, 75005, Paris, France}
\author{Tim de Jongh}
\thanks{Present address: JILA, National Institute of Standards and Technology, and Department of Physics, University of Colorado, Boulder, CO 80309, USA}
\affiliation{Laboratoire Kastler Brossel, ENS-Universit\'{e} PSL, CNRS, Sorbonne Universit\'{e}, Coll\`{e}ge de France, 24 rue Lhomond, 75005, Paris, France}
\author{Bruno Peaudecerf}
\affiliation{Laboratoire Collisions Agr\'egats R\'eactivit\'e, UMR 5589, FERMI, UT3, Universit\'e de Toulouse, CNRS, 118 Route de Narbonne, 31062, Toulouse CEDEX 09, France}
\author{Shiwei Zhang}
\thanks{Correspondence to be addressed to: \href{mailto:tarik.yefsah@lkb.ens.fr}{tarik.yefsah@lkb.ens.fr}, \href{mailto:szhang@flatironinstitute.org}{szhang@flatironinstitute.org}, and \href{mailto:heyuanyao@nwu.edu.cn}{heyuanyao@nwu.edu.cn}}\affiliation{Center for Computational Quantum Physics, Flatiron Institute, 162 5th Avenue, New York, New York 10010, USA}
\author{Tarik Yefsah}
\thanks{Correspondence to be addressed to: \href{mailto:tarik.yefsah@lkb.ens.fr}{tarik.yefsah@lkb.ens.fr}, \href{mailto:szhang@flatironinstitute.org}{szhang@flatironinstitute.org}, and \href{mailto:heyuanyao@nwu.edu.cn}{heyuanyao@nwu.edu.cn}}
\affiliation{Laboratoire Kastler Brossel, ENS-Universit\'{e} PSL, CNRS, Sorbonne Universit\'{e}, Coll\`{e}ge de France, 24 rue Lhomond, 75005, Paris, France}

\date{April 3, 2025. Revised: October 20, 2025}

\begin{abstract}
\quad In this work we explore two-dimensional attractive Fermi gases at the microscopic level by probing spatial charge and spin correlations in situ. Using atom-resolved continuum quantum gas microscopy, we directly observe fermion pairing and study the evolution of two- and three-point correlation functions as inter-spin attraction is increased. The precision of our measurement allows us to reveal nonlocal anticorrelations in the pair correlation function, fundamentally forbidden by the mean-field result based on Bardeen-Cooper-Schrieffer (BCS) theory but whose existence we confirm in exact auxiliary-field quantum Monte Carlo calculations. We demonstrate that the BCS prediction is critically deficient not only in the superfluid crossover regime but also deep in the weakly attractive side. Guided by our measurements, we find a remarkable relation between two- and three-point correlations that establishes the dominant role of pair-correlations. Finally, leveraging local single-pair losses, we independently characterize the short-range behaviour of pair correlations, via the measurement of Tan's Contact, and find excellent agreement with numerical predictions. Our measurements provide an unprecedented microscopic view into two-dimensional Fermi gases and constitute a paradigm shift for future studies of strongly-correlated fermionic matter in the continuum.
\end{abstract}

\maketitle

\nolinenumbers

Ultracold Fermi gases provide a powerful testbed for the exploration of two-dimensional (2D) fermionic matter, with a high level of control and exquisite probing capabilities~\cite{gross2021, daley2022}. The two-component (spin $\up$ and spin $\down$) Fermi gas with attractive contact interactions constitutes a paradigmatic model of many-body physics, with a rich phenomenology and deep connections with condensed matter physics.

\begin{figure*}[ht!]
    \centering
    \includegraphics[width=\textwidth]{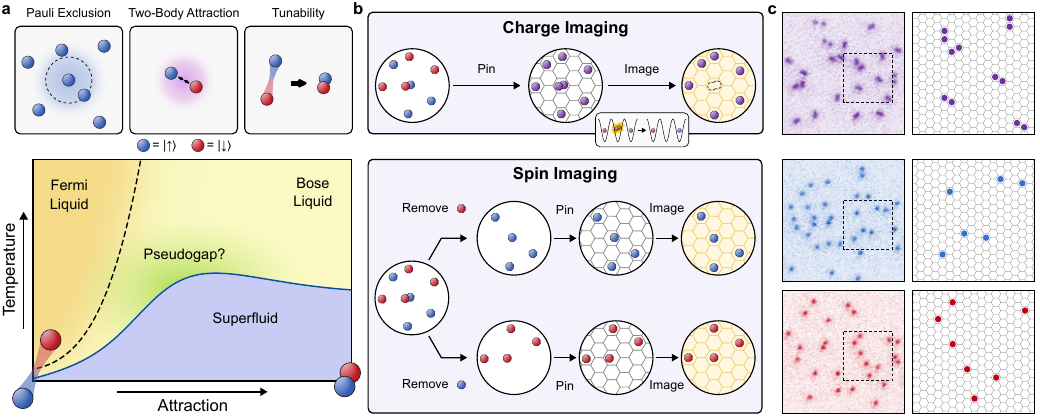}
    \caption{\textbf{Single-Charge and Single-Spin imaging of interacting 2D Fermi gases.} \textbf{(a)} Phase diagram of the spin-balanced 2D attractive Fermi gas across the BEC-BCS crossover. At sufficiently low temperatures, the system is superfluid for any non-zero attraction. At higher temperatures, it exhibits well-understood Fermi and Bose liquid behaviours, but the state right above the normal-to-superfluid temperature is believed to display a pseudogap behaviour below a certain threshold (black dashed line). \textbf{(b)} The spin-$1/2$ mixture is imaged via atom-resolved quantum gas microscopy giving access to the total density (charge), the spin--$\up$ component, and the spin--$\down$ component. Single-spin images are obtained after removal of the other spin component using a resonant light pulse. Sites occupied by two atoms upon pinning appear empty after imaging due to light-assisted collisions, which we use as an independent probe of short-range correlations. \textbf{(c)} Raw single-charge (top) and single-spin (middle and bottom) experimental images along with their processing. In the shown charge image, fermions are seen to organise by pairs.}
    \label{fig:fig1}
\end{figure*}

Crucially, the inter-spin attraction can be varied experimentally by means of a Feshbach resonance from very weak to arbitrarily large, where two opposite spins form a bosonic dimer, thus realizing a unique situation where the quantum statistics of the system can be dynamically tuned. At sufficiently low temperature, an ensemble with balanced spin-populations can be continuously driven from the state of a Bardeen-Cooper-Schrieffer (BCS) superfluid, with non-local fermion pairs, to a Bose-Einstein Condensate (BEC) of tightly bound dimers~\cite{levinsen2015,ries2015,sobirey2021}, remaining in the superfluid state for any interaction strength (see Fig.~\ref{fig:fig1}a). At higher temperature, the normal phase is well understood with the system evolving from a Fermi to a Bose liquid behaviour when the attraction is increased, but at intermediate temperature and sufficiently strong interaction a pseudogap regime may exist right above the normal-to-superfluid transition line, where precursor pairs are formed in the absence of superfluidity~\cite{feld2011,bauer2014,murthy2018}. In quasi-2D superconductors, the microscopic origin of the pseudogap remains largely debated~\cite{fischer2007}. 

In addition, the reduced dimensionality raises fundamental questions. In two-dimensions, the Mermin-Wagner theorem precludes the existence of true long-range order in the thermodynamic limit when interactions are short-ranged~\cite{mermin1966}, and the transition to the superfluid state is of the Berezinskii-Kosterliz-Thouless (BKT) type~\cite{berezinsky1971,kosterlitz1973}. For finite-size samples, the possibility of full coherence over the system leads to a subtle interplay with the BKT transition~\cite{hadzibabic2011}, with the existence of a significant condensed fraction even above the transition. Dimensionality also has profound effects on the symmetries of the system, which can evolve from being scale invariant~\cite{hung2011,yefsah2011} to displaying a quantum anomaly~\cite{holten2018,murthy2019} depending on the interaction strength. Finally, spin-imbalanced 2D Fermi gases are believed to be prime candidates for hosting the elusive Fulde-Ferrell-Larkin-Ovchinnikov phase~\cite{fulde1964,larkin1965,vitali2022}.

Continuum quantum gas experiments have had great success in investigating strongly-interacting 2D fermions, with a wide range of measurements addressing their thermodynamics~\cite{dyke2011,frohlich2011,makhalov2014,fenech2016,boettcher2016}, transport~\cite{bohlen2020,sobirey2021}, coherence~\cite{ries2015,murthy2015,murthy2019}, or spectral~\cite{sommer2012,frohlich2012,murthy2018,sobirey2022} properties. For almost two decades, however, the available imaging techniques have limited the ability to acquire quantitative knowledge at the microscopic scale, i.e., resolving their spatial organisation at length scales below the inter-particle spacing. Recent progress in that direction was achieved for magnified few fermion systems~\cite{brandstetter2025,brandstetter2024}.

In this work, we probe strongly-interacting 2D Fermi gases at low temperatures via atom-resolved in-situ imaging. By measuring spatial charge and spin correlations over a wide range of interaction strengths, we reveal the intricate microscopic behaviour of this strongly-correlated system. While strongly-interacting fermionic ensembles are in general extremely challenging to tackle theoretically, the case of equal spin-populations represents a rare example for which numerically exact calculations can be performed~\cite{shi2015,vitali2017,he2019,he2022}. We take advantage of this to perform a high-precision comparison of our measurements with cutting-edge auxiliary-field quantum Monte Carlo (AFQMC) calculations. We also contrast our results with BCS theory, charting the regimes of its breakdown for various microscopic observables. Our work demonstrates capabilities to directly and precisely measure spatial correlations in a strongly-correlated system, which will readily apply to situations where accurate computations are significantly more challenging, such as Fermi gases where spin populations are unequal, as well as where interactions are repulsive or long-range.

Our experiment starts with a two-component spin mixture of fermionic $^6$Li atoms ($\ket{\up}$ and $\ket{\down}$) with equal spin populations. Atoms are confined to a single plane using a strong laser-induced trap in the vertical $z$--direction, with frequency $\omega_z/2\pi=1.125(50)$\,kHz corresponding to a temperature $T_z=\hbar\omega_z/k_{\rm B}=54(2)$\,nK, with $\hbar$ the reduced Planck constant and $k_{\rm B}$ the Boltzmann constant. This temperature scale is much larger than the absolute temperature $T\lesssim5$\,nK, and at least on the order of  the Fermi temperature $T_{\rm F}=\pi n\hbar^2/mk_{\rm B}$ of our samples, $n$ being the total density and $m$ the atomic mass. This corresponds to the interesting regime where the system can significantly depart from the purely 2D regime only in the presence of strong interactions, allowing the population of higher $z$--levels. This quasi-2D geometry and the finite size of our cloud favours an increase of the critical temperature on the order of $T_{\rm c}\sim0.15\,T_{\rm F}$ in the crossover \cite{fischer2014,ries2015}, such that most of our samples are in the superfluid regime with reduced temperatures as low as $T/T_{\rm F}\approx0.08$.

\begin{figure*}[t]
    \centering
    \includegraphics[width=\textwidth]{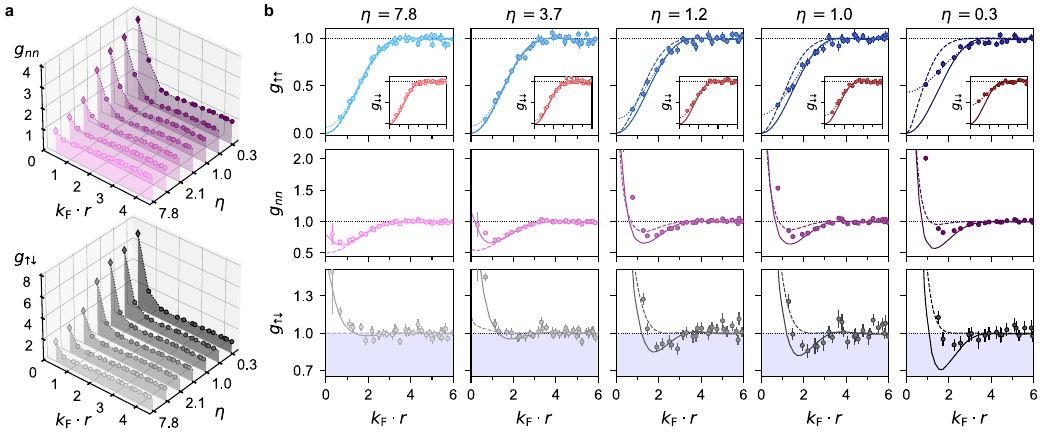}
    \caption{\textbf{ Two-point charge and spin correlations.} \textbf{(a)} 
Charge and pair correlations as a function of the interaction parameter $\eta$ (the values in the labels are rounded for readability). The diamonds at short distance are obtained via pair-loss measurements (see text). The upshoot at short range represents a direct observation of fermion pairing in real space. \textbf{(b)} Close-up of correlations for selected values of $\eta$. Top row: Equal-spin correlations. Middle row: Charge correlations. Bottom row: Pair correlations. Dashed lines: mean-field BCS theory. Solid lines: AFQMC calculations for the 2D gas at our experimental values of $\eta$ and reduced temperature $T/T_{\rm F}$. Dotted lines (top row): two-parameter fit of $g_{\sigma\sigma}$ (see \cite{Note1}). For all interactions the measured correlations display clear deviations from the BCS prediction. Furthermore, inter-spin correlations display a dip where $g_{\uparrow\downarrow}<1$, in violation of BCS theory, which is also observed in the AFQMC results. In the weakly interacting regime we find excellent agreement between our data and AFQMC. In the crossover region interactions induce the occupation of excited $z$-levels, leading to a reduced contrast in $g_{\up\up}$ and expected differences with AFQMC in 2D.}
    \label{fig:fig2}
\end{figure*}

Inter-spin interactions are characterized by the 2D scattering length
\beq
a=\ell_z\sqrt{\pi/B}\exp(-\sqrt{\pi/2}\,\ell_z/a_{\rm 3D}),
\eeq reflecting the combined effect of the inter-spin scattering length in three-dimensional space $a_{\rm 3D}$ and the vertical harmonic oscillator length $\ell_z=\sqrt{\hbar/m\omega_z}$, with $B\approx0.905$  \cite{petrov2001,pricoupenko2007, levinsen2015}. At the many-body level, the relevant interaction parameter is $\eta=\log(k_{\rm F}a)$, where $k_{\rm F}=\sqrt{2\pi n}$ denotes the Fermi wavenumber. The BCS and BEC regimes correspond respectively to $\eta\gg1$ and $\eta\ll-1$, where the pair size is much greater than the inter-particle spacing in the former case, and much smaller in the latter. 

We probe the system using continuum quantum gas microscopy, 
which consists in freezing the motion of the atoms initially evolving in continuous space, before imaging their positions. This is achieved by suddenly turning on a pinning optical lattice, and subsequently exposing the atoms to cooling light, which drives their fluorescence while maintaining them in individual lattice wells \cite{verstraten2025,jongh2024}. Prior to pinning, we either remove one of the spin components, which results in images of single-spin spatial distributions, or we directly image the total density, corresponding to single-charge imaging, as depicted in Fig.~\ref{fig:fig1}b. Charge imaging does not allow the detection of doubly occupied sites \footnote{Additional details can be found in the Supplementary Materials.} but owing to the extreme diluteness of our clouds, this does not introduce any detrimental effect, and we later exploit this feature to measure the short-range behaviour of the pair correlations. By acquiring charge, spin-$\up$, and spin-$\down$ images from $\sim 750$ separate experimental runs each, we obtain all relevant two-point correlation functions:

\beqa
g_{nn}(r)&=&\frac{\brak{n(0)n(r)}}{\brak{n(0)}\brak{n(r)}},\nonumber\\
g_{\sigma\sigma}(r)&=&\frac{\brak{n_\sigma (0)n_\sigma (r)}}{\brak{n_\sigma (0)}\brak{n_\sigma (r)}},\quad\sigma=\up,\down\nonumber\\
g_{\up\down}(r)&=&\frac{\brak{n_\up(0)n_\down(r)}}{\brak{n_\up(0)}\brak{n_\down(r)}},
\label{eq:g2}
\eeqa with $r\neq0$. The functions $g_{nn}$, $g_{\up\up}$, and $g_{\down\down}$ are directly measured from the data, while $g_{\up\down}=g_{\down\up}$ is retrieved via the relation:
\beq
g_{nn}(r)=\frac{1}{2}g_{\up\up}(r)+\frac{1}{2}g_{\up\down}(r), 
\label{eq:gnn}
 \eeq
valid in the spin-balanced case, where $g_{\up\up}=g_{\down\down}$.

In Fig.~\ref{fig:fig2}a, we show an overview of the measured spatial charge and pair correlations for seven interaction strengths $\eta =  7.77^{+0.38}_{-0.34}$, $3.70^{+0.21}_{-0.18}$, $2.08^{+0.13}_{-0.13}$, $1.21^{+0.10}_{-0.09}$, $0.98^{+0.09}_{-0.09}$, $0.66^{+0.06}_{-0.06}$,  and $0.28^{+0.06}_{-0.08}$, at reduced temperatures $T/T_{\rm{F}} = 0.18(2)$,  $0.16(1)$, $0.11(1)$,  $0.09(1)$,  $0.09(1)$,  $0.08(1)$,  and $0.08(1)$, respectively. The effect of interactions is immediately visible in the upshoot of $g_{nn}$ at short distance that increases with the attraction. This signals the formation of increasingly tight $\up\down$--pairs as one moves from the BCS towards the BEC side. A close up of these correlations for representative values of $\eta$ is shown in Fig. \ref{fig:fig2}b (see also extended data in \cite{Note1}), and reveals their highly non-local nature with prominent anti-correlations at a length scale $\sim 2k_{\rm F}^{-1}$. In the absence of interaction, the spin-$\up$ and spin-$\down$ components are uncorrelated, hence $g_{\up\down}(r)=1$, and $g_{nn}(r)=(g_{\up\up}(r)+1)/2$ tends to 1/2 at $r=0$.

Our data is compared to both BCS mean-field theory (at $T=0$), and AFQMC calculations performed at the experimentally determined interaction strength $\eta$ and reduced temperature $T/T_{\rm F}$. Remarkably, BCS theory already fails at $\eta\approx7.8$, incorrectly predicting a nearly uncorrelated gas behaviour. Moving towards the center of the crossover, we observe a net violation of BCS theory's constraint $g_{\up\down}\geq1$ \cite{obeso-jureidini2022}. Our measured pair correlations display a marked dip at $k_{\rm F}r\sim2$ reaching values well below 1, which is also present in the AFQMC calculations. This pair-correlation dip signals a correlation beyond independent pairs and is a non-trivial signature in interacting Fermi gases.

\begin{figure*}[t]
    \includegraphics[width=\textwidth]{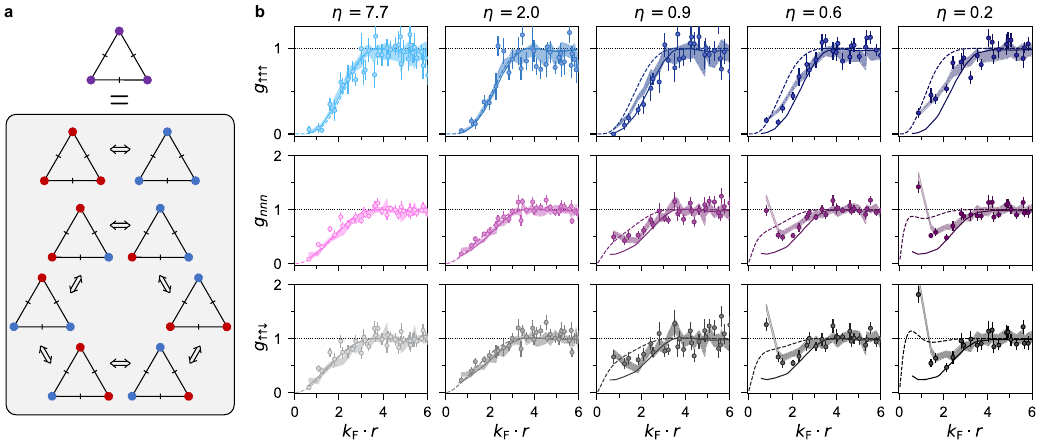}
    \caption{\textbf{Three-point charge and spin correlations.} \textbf{(a)} On equilateral triangles and for balanced spin-populations there are enough symmetries to extract all relevant three-point correlation functions from the single-spin and single-charge images. \textbf{(b)} Measured three-point correlations on equilateral triangles for various values of the interaction parameter $\eta$ (rounded for readability). The small difference in the values of $\eta$ with respect to Fig.~\ref{fig:fig2} is due to the use of a slightly larger region of the cloud to measure $g_3$ (see \cite{Note1}). Dashed lines: zero-temperature BCS prediction. Solid lines: AFQMC calculations at the same reduced temperatures as in the experiments, showing very good agreement for $\eta\lesssim2$, as well as in the crossover when $k_{\rm F}r\gtrsim2$. The observed discrepancies at short distance are expected due to interaction-induced population of excited $z$--levels, leading to an offset in $g_{\up\up\up}$ and an enhanced rise of $g_{nnn}$ and $g_{\uparrow\uparrow\downarrow}$. The shaded areas are the experimental results obtained by applying Eqs.~(\ref{eq:g3wick1}--\ref{eq:g3wick3}) to the measured two-point correlations, showing very good agreement for all interactions.}
    \label{fig:fig3}
\end{figure*}

Quantitatively, excellent agreement is found between experiments and AFQMC calculations for $\eta\approx7.8$ and 3.7 when the samples are well in the 2D regime \cite{Note1}. For smaller values of $\eta$, which are closer to the crossover, the large interaction energy allows fermions to occupy higher motional $z$-levels of the vertical confinement, not taken into account in the simulations. The exact role of the third dimension, which relates to important questions in the context of quasi-2D superconductivity \cite{fischer2007} and is a source of interesting theoretical opportunities \cite{zujev2014,fontenele2024}, will be the subject of future work. The occupation of higher $z$-levels is evidenced in the measured $g_{\up\up}$ correlations (Fig.~\ref{fig:fig2}b, first row) by an offset at $r=0$ whose value is a measure of the excited states population \cite{jongh2024}. In the strictly 2D case, our AFQMC calculations show that $g_{\up\up}$ is indistinguishable from the one of an ideal Fermi gas at the corresponding temperature \cite{Note1} for all interactions values considered here. This is in stark contrast to the result of BCS theory (see Fig.~\ref{fig:fig2}b, first row), which predicts that the Fermi hole shrinks significantly near the crossover. Instead, we find that all the way from the far BCS side to the middle of the crossover near $\eta=0$, particles organize themselves in a remarkable manner where each spin-component behaves as an ideal Fermi gas, while ensuring strong inter-spin correlations. This raises the fundamental question of whether higher order correlations contain any information that is not already contained in the two-point equal-spin and pair-correlations.

To address this question we measure the charge, equal-spin, and inter-spin three-point correlations functions from the same samples. For simplicity, we focus on correlations between three equidistant fermions, which only depend on a single spatial parameter $k_{\rm F}r$, with $r$ the distance between each pair. This symmetry, which is native to our triangular pinning lattice~\cite{Note1}, has the crucial advantage to make all six inter-spin correlations equivalent (see Fig.~\ref{fig:fig3}a), and one can show that:

\beq
g_{nnn}(r)=\frac{1}{4}g_{\up\up\up}(r)+\frac{3}{4}g_{\up\up\down}(r), 
\label{eq:gnnn}
\eeq where $g_{nnn}$, $g_{\up\up\up}$, and $g_{\up\up\down}$ are defined according to the same convention as their two-point counterparts in Eq.~(\ref{eq:g2}). As previously, $g_{nnn}$ and $g_{\up\up\up}$ are directly measured from the images, while  $g_{\up\up\down}$ is retrieved via Eq.~(\ref{eq:gnnn}), which is only possible owing to the symmetries of the equilateral triangle. Our results are reported in Fig.~\ref{fig:fig3}b for $\eta=7.72^{+0.39}_{-0.35}$, $2.04^{+0.14}_{-0.15}$, $0.94^{+0.13}_{-0.12}$, $0.61^{+0.11}_{-0.10}$, and $0.22^{+0.10}_{-0.11}$, at reduced temperatures $T/T_{\rm{F}} =0.18(2)$, $0.11(1)$, $0.09(1)$, $0.08(1)$, and $0.08(1)$, respectively. In contrast to the case of two-point correlations, the prediction of BCS theory agrees well with our data and the AFQMC calculations up to $\eta\approx2$, and only starts failing in the crossover, displaying the wrong trend for $k_{\rm F}r\lesssim3$. The observed rise of the measured charge and inter-spin correlations at short distance strikingly contradicts the drop to zero of the BCS result. We expect the presence of interaction-induced occupation of $z$-levels to enhance correlations at short-distance, but the possibility of a true divergence already present in strictly 2D systems cannot be excluded. While one may argue that the proximity of a pair of equal spins should necessarily result in the vanishing of these correlations at short distance, this argument was invalidated in 3D, where it was shown analytically that the inter-spin three-point correlation diverges at short distance \cite{werner2024}. The question is not answered in 2D. AFQMC calculations also do not capture the rapid rise of the measured $g_{nnn}$ and $g_{\up\up\down}$ in the crossover, which we attribute to the presence of interaction-induced occupation of $z$-levels, but the agreement is nonetheless very good down to $k_{\rm F}r\sim2$, and the upturn displayed by the calculations might be the precursor of a true divergence. Our observations call for analytical studies of the short-range behaviour of three-point correlations in 2D Fermi gases.

To evaluate the extent of the dependence of three-point correlation functions on the two-point ones, we compute the following quantities using the measured correlations:

\beqa
\tilde{g}_{\up\up\up}(r)&=&3g_{\up\up}(r)+2[1-g_{\up\up}(r)]^{3/2}-2\label{eq:g3wick1}\\
\tilde{g}_{nnn}(r)&=&\frac{3}{2}\left[g_{\up\up}(r)+g_{\up\down}(r)\right]\nonumber\\
&&+\quad\frac{3}{2}\left[g_{\up\up}(r)-g_{\up\down}(r)\right]\sqrt{1-g_{\up\up}(r)}\nonumber\\
&&+\quad2\left[1-g_{\up\up}(r)\right]^{3/2}-2\label{eq:g3wick2}\\
\tilde{g}_{\up\up\down}(r)&=&g_{\up\up}(r)+2g_{\up\down}(r) \nonumber \\
&&+2\left[ g_{\up\up}(r) - g_{\up\down}(r) \right]\sqrt{1-g_{\up\up}(r)} \nonumber \\
&&+2\left[1-g_{\up\up}(r)\right]^{3/2}-2.
\label{eq:g3wick3}
\eeqa These relations are conjectured based on the Wick theorem, which applies to non-interacting or weakly interacting Fermi gases, but is expected to fail in the strongly-interacting regime. The obtained results are reported in Fig.~\ref{fig:fig3}b as shaded areas. Remarkably, measured three-point correlations $g_{\up\up\up}$, $g_{nnn}$, and $g_{\up\up\down}$ agree to a high degree with the $\tilde{g}_{\up\up\up}(r)$, $\tilde{g}_{nnn}(r)$ and $\tilde{g}_{\up\up\down}(r)$ obtained from the measured two-point correlations via Eqs.\,(\ref{eq:g3wick1}--\ref{eq:g3wick3}). This observation should not, however, be interpreted as the Wick theorem being valid here, and we have verified using the AFQMC calculations that Wick relations involving the lowest order correlations fail in the crossover \cite{Note1}. Instead this result signals that three-point correlations do not contain additional information compared to the two-point ones. A complete interpretation of the strongly-correlated nature of the system requires investigating even higher order correlations, and specifically the role of pair-pair correlations. These open questions can be readily addressed with our combined experimental and computational capabilities.

\begin{figure}[t]
    \centering
    \includegraphics[width=\columnwidth]{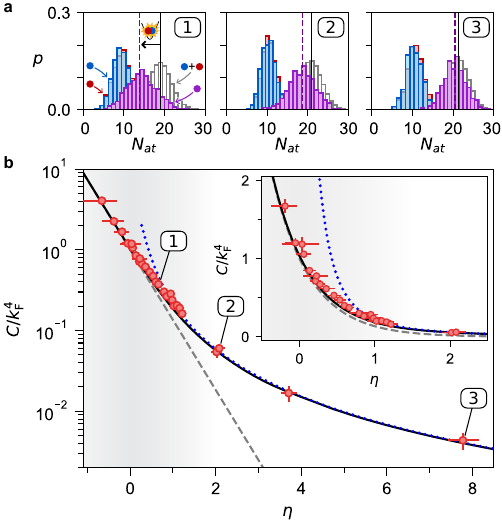}
    \caption{\textbf{Contact density.} \textbf{(a)} Distribution of the atom number for spin $\uparrow$ (red), spin $\downarrow$ (blue) and both spins (purple) in a central subregion of the cloud for $\eta\approx 0.7$, $2.1$, and $7.8$. The solid black line is the sum of the average atom number in state $\uparrow$ and in state $\downarrow$. The dashed purple line is the average atom number in images where no atoms are removed. In grey, we show the expected histogram for $N_{\uparrow} + N_{\downarrow}$ in the absence of light-assisted collisions. For increasing attraction, losses due to double occupancies become larger and yield a direct measure of the probability to find a $\up\down$--pair on the same site. \textbf{(b)} Contact density as a function of $\eta$. The contact density is measured via losses in different quasi-homogeneous subregions of the cloud which correspond to different values of the interaction parameter $\eta$. The grey dashed line is the BCS mean-field prediction $1/(k_{\rm F}a)^2$. The blue dotted line is obtained from Fermi-liquid theory \cite{engelbrecht1992}. The solid line is the result of AFQMC calculations at $T\, =\, 0$ \cite{shi2015}. Inset: linear scale contact density in the crossover.}
    \label{fig:fig4}
\end{figure}

We now turn to the short-range behaviour of the pair-correlation function, which reads \cite{werner2012}:
\beq
g_{\uparrow \downarrow}(r) \underset{r\to0}{=}4\,\frac{C}{k_{\rm F}^4} \log^2\left(\alpha\frac{r}{a}\right),
\label{eq:g2_C}
\eeq where $\alpha=e^\gamma/2$ and $\gamma$ is Euler's constant. While this scaling is fully determined by two-body physics, the value of the coefficient $C$, the contact, is set by the many-body behaviour of the system and depends on the interaction strength $\eta$. The contact represents an important overarching quantity, connecting the microscopic and macroscopic behaviour of the system \cite{tan2008a}. Eq.\,\ref{eq:g2_C} shows that the contact is linked to the probability of finding two opposite spins within a probe volume whose characteristic length is smaller than the relevant length scales of the system, $k_{\rm F}^{-1}$ and $a$. In our experiment, the lattice sites after projection constitute natural probe volumes to measure pair occupations. Owing to the extreme diluteness of our samples, the occupation of a lattice site by three or more particles is strongly suppressed, and the only relevant events are the single and double occupancies. Double occupancies are dominated by $\up\down$--pairs, since the probability for equal-spin pairs is strongly reduced by Pauli's exclusion principle. To count the number of $\up\down$--pairs per lattice site, we exploit light-induced losses that occur during imaging when two atoms occupy the same site, a known detrimental effect of quantum gas microscopy which we are able to turn here into a powerful probe of the contact parameter. By comparing the average atom numbers in single-spin and charge images respectively (see Fig.\,\ref{fig:fig4}a), we infer the the number of lost atoms, and therefore the probability to find a $\up\down$-pair per lattice site, taking into account the residual contribution of the equal-spin double occupancies \cite{Note1}. The resulting contact measurement as a function of interaction strength is shown in Fig.\,\ref{fig:fig4}, which we compare to zero-temperature AFQMC calculations \cite{shi2015}, finding excellent agreement over three orders of magnitude. The comparison obtained here sets a new standard of cross-validation between experimental measurements and high precision computations for microscopic phenomena in continuum systems.

We have probed an interacting Fermi gas with tuneable inter-spin attraction via atom-resolved in-situ imaging, directly observing fermion pairing and the spatial form of all relevant two- and three-point correlation functions. Our continuum quantum gas microscopy technique allowed us to probe these correlations at length scales below the inter-particle spacing. We obtained a characterisation of the system at the microscopic level with unprecedented precision, revealing critical deviations from BCS theory in both the superfluid crossover regime and the weakly attractive side, and we found a relationship between two- and three-point correlations that underscores the dominant role of pair correlations. We compared our measurements with numerically exact AFQMC computations, confirming their key qualitative features and finding excellent quantitative agreement when the systems are directly comparable. Our work, which combines atom-resolved experimental measurements and AFQMC calculations, can be readily extended to the study of fourth or higher order correlations and the finite temperature regime just above the normal-to-superfluid transition and beyond. A crucial direction is the exploration of strongly-interacting fermions with unequal spin-populations or repulsive interactions, where accurate computations are significantly more challenging. The experimental advances demonstrated here add a powerful dimension to the study of correlated quantum systems in the continuum and offer a unique opportunity to obtain a profound understanding of their microscopic inner workings.\\ 

\textit{Note} -- During the completion of this work, we became aware of other work on quantum gas microscopy in the continuum \cite{yao2024,xiang2024}.\\ 

\textbf{Acknowledgements:} We thank Yvan Castin, Lawrence Cheuk, Axel Pelster and his group, Carlos S{\'a} De Melo, Kris Van Houcke, F\'elix Werner, and Martin Zwierlein for insightful discussions, and F\'elix Werner for a critical reading of the manuscript. T.Y is grateful to Antoine Heidmann for his crucial support as head of Laboratoire Kastler Brossel. This work has been supported by Agence Nationale de la Recherche (Grant No. ANR-21-CE30-0021), CNRS (Tremplin@INP 2020), and R{\'e}gion Ile-de-France in the framework of DIM SIRTEQ and DIM QuanTiP. The Flatiron Institute is a division of the Simons Foundation. Y-Y. H. was supported by the National Natural Science Foundation of China (under Grants No. 12247103 and No. 12204377), and the Youth Innovation Team of Shaanxi Universities.\\

\textbf{Author contributions:} C.D., M.D. and J.V. performed the experiment. C.D., M.D., J.V., T.d.J, and B.P, all contributed to the analysis of the data. Y.Y.H and S.Z. carried out the AFQMC calculations. T.Y. planned and supervised the study. All authors contributed to the interpretation of the results and to the writing of the manuscript.\\

\textbf{Competing interests:} The authors declare no competing interests.

\newpage

\nolinenumbers

\bibliographystyle{apsrev4-2}

\newpage

\renewcommand\thefigure{S\arabic{figure}}
\renewcommand\theHfigure{S\arabic{figure}}
\setcounter{figure}{0} 

\renewcommand\theequation{S\arabic{equation}}
\renewcommand\theHequation{S\arabic{equation}}
\setcounter{equation}{0} 

\noindent
\pagebreak

\section*{Supplementary Materials}\label{sec:Methods}

\nolinenumbers

\subsection*{Experimental Sequence}

The experimental setup was described in detail in \cite{jin2024,verstraten2025}. The experiment starts with a balanced mixture of $^6$Li atoms in the first and third lowest-energy hyperfine states denoted $\ket{1}\equiv\ket{\up}$ and $\ket{3}\equiv\ket{\down}$ in a 'light sheet' dipole trap which provides weak (resp. strong) harmonic confinement in the $xy$-plane (resp. transverse $z$-direction). We perform evaporative cooling in the light sheet at 690\,G close to the $\ket{1}-\ket{3}$ Feshbach resonance down to a trap depth of $\sim$\,80\,nK over a duration of $\sim$4\,s. The trap is then adiabatically ramped up to a depth of $\sim$ 215\,nK over 100\,ms, corresponding to a transverse frequency $\omega_z = 2\pi\, \times \,1125(50)$\,Hz. The Feshbach field is swept in 50\,ms to tune the interactions. Subsequently the gas is held for 500\,ms to ensure thermalisation.
\\

Afterwards, atoms in state $\ket{1}$ or $\ket{3}$ can optionally be removed by using a strong resonant light pulse of duration $t_{\rm blast}\sim20\,\mu$s ($I\sim I_{\rm sat}$) propagating along the $z$-direction. Towards the end of this blasting pulse, we ramp up the pinning lattice in $t_{\rm pin} \simeq 5$\,$\mu$s. The pinning lattice is created by the self-interference of a red-detuned 1064\,nm laser beam. Three arms cross at 120$^\circ$ angles in the horizontal plane which results in a triangular lattice geometry with spacing $a_L\,=\,709\,$nm. The total duration of this procedure, $t_{\rm blast} + t_{\rm pin} \leq 25\,\mu$s is kept smaller than all other physical timescales in the system, ensuring that the information on the density is preserved during the pinning process. The magnetic field is then ramped down to 0\,G in 30\,ms, followed by Raman sideband cooling. We collect the fluorescence signal using a high NA objective and an EMCCD camera. The resulting  images are analyzed with a high-fidelity neural network to obtain the atomic positions. For each experimental run, we take two successive images of the cloud with a 600\,ms exposure time and a 250\,ms interval between each image. This allows us to measure the fraction of atoms that remain pinned within their lattice site, which is above 99\,\% for all preparations.
\\

For this work we calibrated all values of the Feshbach field used via radio-frequency spectroscopy. The value of the transverse trap frequency $\omega_z$ used to calculate the 2D scattering length was measured via parametric modulation. Each preparation corresponds to between 1500 and 3000 experimental images, evenly split between blasting state $\ket{1}$, blasting $\ket{3}$ and not blasting. In practice we alternate between these three different imaging routines.

\subsection*{Spin balance}

\begin{figure}[!t]
    \centering
    \includegraphics[width=\columnwidth]{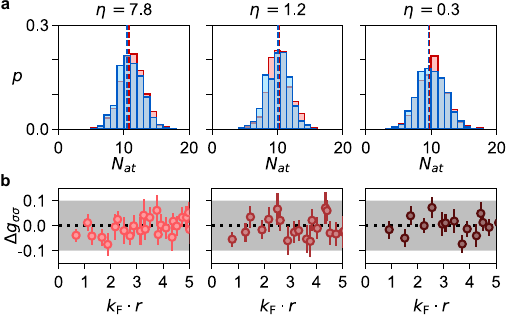}
	\caption{\textbf{Spin balance}.\textbf{(a)} Histogram of the atom number measured in a central region of the cloud when removing $\ket{1}$ (red) and removing $\ket{3}$ (blue) for 3 different interaction strengths. \textbf{(b)} Absolute difference $\Delta g_{\sigma\sigma} = g_{\uparrow\uparrow} - g_{\downarrow\downarrow}$ of the corresponding correlation functions.}
    \label{fig:Spin_balance}
\end{figure}

Measuring $\ket{1}$ and $\ket{3}$ independently allows us to check that the spin mixture is balanced to better than $1\%$. While we observe on average a $\sim$3.5\,\% excess when measuring state $\ket{3}$ compared to state $\ket{1}$, we attribute a large fraction of this difference to the fact that no closed transition is available from state $\ket{1}$, unlike state $\ket{3}$. An excited state $\ket{1'}$ can therefore decay into a dark state during the blasting process, resulting in a lower removal efficiency of state $\ket{1}$ atoms. Around 700\,G, this occurs once every $\sim$360 emission cycles. We estimate that, on average, an atom needs to absorb $\sim$10 photons to leave the trap, which means the probability for an atom in state $\ket{1}$ to absorb fewer than 10 photons before decaying into a dark state is $2.7\,\%$ which accounts for most of the observed excess atoms when removing state $\ket{1}$. In Fig.~\ref{fig:Spin_balance}, we show  the atom numbers when removing either $\ket{1}$ or $\ket{3}$ as well as the difference of the corresponding correlation functions $g_{\sigma\sigma}$. This imperfect removal leads to an artificially reduced contrast of $g_{\down\down}$ at short distances, and we therefore use $g_{\uparrow\uparrow}$ throughout the paper.

\subsection*{Thermometry}

\begin{figure}[!t]
    \centering
    \includegraphics[width = \columnwidth]{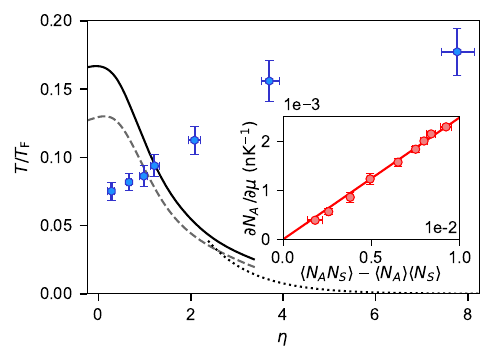}
	\caption{\textbf{Fluctuation thermometry}. Reduced temperatures of our samples (near the centre of the clouds) obtained via fluctuation-dissipation thermometry. The solid black line corresponds to the critical temperature for superfluidity in a finite-size 2D system, while the dashed grey line corresponds to the critical
temperature for the BKT transition at the thermodynamic limit (from \cite{he2022}). The dotted line is the expected critical temperature in the BCS limit \cite{petrov2003}. Inset: Example of the relation between compressibility and atom number covariance obtained for our strongly-interacting Fermi gases ($\eta\approx0.3$ near the center of the cloud) and its linear fit. The slope of the fit gives the inverse temperature of the sample.}
    \label{fig:Thermometry_vb}
\end{figure}

To measure the temperature we combine our recently introduced correlation-based thermometry method \cite{dixmerias2025} with a model-independent approach based on the fluctuation-dissipation theorem \cite{zhou2011,hartke2020}. In a system $S$ whose thermodynamic properties can be described within the grand canonical ensemble, the covariance between the atom number inside a small probe volume $A$ and the atom number inside the full system is related to the absolute temperature \cite{zhou2011}:
\beq
k_B T  \frac{\partial \langle N_A \rangle}{\partial \mu}  = \langle N_A N_S \rangle - \langle N_A \rangle \langle N_S \rangle
\eeq
In the case of a mixture this relationship holds and should be applied independently on each component. Within the local density approximation (LDA), if the potential landscape is known, the compressibility can be computed directly from the measured density profile and the inverse temperature is obtained from a linear fit to the curve \mbox{$-\frac{\partial N_A}{\partial V} = f(\langle N_A N_S \rangle - \langle N_A \rangle \langle N_S \rangle)$, as shown in Fig.~\ref{fig:Thermometry_vb}.}
\\
We precisely calibrate our trapping potential by producing a non-interacting Fermi Gas and measuring its temperature using a correlation-based thermometry method \cite{dixmerias2025}. The potential is then reconstructed by inverting the known equation of state of a quasi-2D non interacting Fermi gas.
We apply this thermometry method on all our data and find absolute temperatures in the range $4-6$\,nK corresponding to reduced temperatures in the range $T/T_{\rm{F}}\sim0.08$ to $0.18$, where $T_{\rm{F}}$ is the average Fermi temperature in a central region of the cloud. $A$ is taken to be a single lattice site while $S$ comprises all sites within a circle of radius 5\,$a_L$ around $A$. $S$ is thus slightly larger than the measured correlation length of our samples while remaining small compared to the full size of the cloud, ensuring that it is well described by the grand-canonical ensemble. By measuring the temperature for each spin component individually we obtain similar results, proving that the two spin states are in thermal equilibrium with one another.

\subsection*{Computation of correlation functions}

We compute correlation functions using the positions of the detected atoms in the pinning lattice. We restrict the analysis to a central quasi-homogeneous region near the center of the cloud where the density is constant within a few percent ($\pm5\%$ for $g_2$, and $\pm7.5\%$ for $g_3$). 
\\
To compute two-point correlations, we loop over all pairs of atoms whose center of mass lies within the central region and compute the probability of finding two atoms at a given distance from each other. Error bars are obtained via bootstrapping and correspond to one standard deviation.

We use a similar algorithm to compute three-point correlators on equilateral triangles. We loop over every pair of atoms in the cloud. Once the sites of the first 2 atoms are known, there are only two possible sites with which to form an equilateral triangle. These are checked for the presence of a third atom and we count the number of atomic triplets which form an equilateral triangle with a given side length and whose center of mass lies within the central region defined above. Similarly to the algorithm for $g_2$, errorbars are obtained via bootstrapping and correspond to one standard deviation.

\subsection*{Effect of lattice discretization}

The imaging process relies on an external lattice to pin and allow fluorescence imaging of the atoms. This leads to an effective imaging resolution given by the shape and size of the elementary Wigner-Seitz (WS) cell of the lattice. In the case of our triangular lattice, the WS cell is a hexagon. For every image we measure occupation numbers in a lattice:
\beq
N_{i,\sigma} = \int_{\textbf{r} \in \rm{WS}_i} n_{\sigma}(\textbf{r})d\textbf{r}
\eeq
Thus, the measured two-point correlator on the lattice, $g_{ij,\sigma\sigma'}$ is:
\beqa
g_{ij,\sigma \sigma '} &=& \frac{\langle N_{i,\sigma} N_{j,\sigma '} \rangle}{\langle N_{i,\sigma} \rangle \langle N_{j,\sigma '} \rangle}\nonumber\\
&=&\frac{\int_{\textbf{r} \in \rm{WS}_i}\int_{\textbf{r}' \in \rm{WS}_j}d\textbf{r}d\textbf{r}'\, \langle n_{\sigma}(\textbf{r})n_{\sigma '}(\textbf{r}')\rangle}{\langle N_{i,\sigma} \rangle \langle N_{j,\sigma '} \rangle}
\eeqa
Provided that $\langle n _{\sigma}(\textbf{r}) \rangle$ is approximately uniform over a WS cell the above expression can be recast in the form:
\beq
g_{ij,\sigma \sigma '} = \int_0^{+\infty}dr\,D_{ij}(r)g_{\sigma \sigma '}(r)
\eeq
where $D_{ij}(r)$ is the probability density of the distance between two points drawn randomly inside cell $i$ and cell $j$. $D_{ij}$ is known analytically for identical and adjacent hexagonal cells and it can be computed numerically for any lattice vector shift $(\delta i, \delta j)$. Thus $g_{ij,\sigma\sigma'}$ is an average of the true $g_{\sigma\sigma'}(r)$, weighted by the frequency of available distances between the two cells considered. In particular this finite resolution leads to a finite value of $g_{ii, \uparrow \downarrow}$, despite the short range divergence of $g_{\uparrow \downarrow}(r)$.
\\

This means that in general \mbox{$g_{ij,\sigma\sigma'}\,\neq\,g_{\sigma\sigma'}(\langle r_{ij} \rangle)$}. To first order, the difference between these two quantities can be linked to the curvature of $g_{\sigma\sigma'}(r)$ over the support of $D_{ij}$. For the correlation functions directly measured in this work, the effect of this finite resolution is usually negligible except on $g_{ij,nn}$ at short distances in the crossover region.

\subsection*{Loss analysis}

In images where both spins states are present we find less atoms than would be expected from the atom numbers measured in single-spin images. These losses are related to the probability of finding an $\uparrow\downarrow$-pair on the same site. We express the measured occupation probability of a site \textit{after imaging} in single-spin (resp. single-charge) images as $p_{\sigma}$ (resp. $p_n$) in terms of the full set of possible events $(i\uparrow,\, j\downarrow)$, corresponding to the presence of $i$ spins $\uparrow$ and $j$ spins $\downarrow$ on a site \textit{before imaging}. Due to light-assisted collisions, only sites with odd occupations appear filled:
\beq
p_{\uparrow} = \sum_{i\ \rm odd}p(i\uparrow) = \sum_{i\ \rm odd}\sum_{j=0}^{+\infty}p(i\uparrow,\,j\downarrow)
\eeq
with a similar expression for $p_{\downarrow}$. For images of both spins:
\beqa
p_{n} &=& \sum_{(i + j)\ \rm odd} p(i\uparrow,\,j\downarrow)\nonumber\\
&=&\left(\sum_{i\ \rm odd}\sum_{j\ \rm even}
+\sum_{i\ \rm even}\sum_{j\ \rm odd}\right) p(i\uparrow,\,j\downarrow)
\eeqa
Losses can then be written as:
\beqa
p_{\uparrow} + p_{\downarrow} - p_n &=& 2\sum_{i\ \rm{odd}}\sum_{j\ \rm odd}p(i\uparrow,\,j\downarrow)\nonumber\\
&=&2\times[p(\uparrow,\,\downarrow) + p(3\uparrow,\, \downarrow) + p(\uparrow,\, 3\downarrow) + \cdots]\nonumber
\eeqa

Owing to the diluteness of our samples and Pauli's exclusion principle the first term largely dominates the rest of the sum. Losses thus provide a direct measure of the probability of finding exactly one spin-$\uparrow$ and one spin-$\downarrow$ on the same site. This probability is related to $g_{ii, \uparrow\downarrow}$ by:
\beqa
g_{ii, \uparrow\downarrow} &=& \frac{\langle N_{i\uparrow}N_{i\downarrow}\rangle}{\langle N_{i\uparrow} \rangle \langle N_{i\downarrow} \rangle}\nonumber\\
&\simeq &\frac{p(\uparrow,\,\downarrow)}{\langle N_{i\uparrow} \rangle \langle N_{i\downarrow} \rangle}
\eeqa
where $\langle N_{\uparrow} \rangle$ and  $\langle N_{\downarrow} \rangle$ are the average occupations on a given site \textit{before imaging}. They differ from the measured occupations probabilities $p_{\uparrow}$ and $p_{\downarrow}$ by the fraction of sites occupied by two atoms of the same spin (the diluteness of our samples and Pauli exclusion lead to negligible triple occupations). The first order correction is given by $\langle N_{\sigma} \rangle \simeq p_{\sigma} \times (1 + 2g_{ii,\sigma\sigma} \times p_{\sigma})$. As we can not measure the value of $g_{ii, \sigma\sigma}$ directly from the experimental images, we estimate its value by extrapolating the measured values of $g_{ij,\sigma\sigma}$ via a two-parameter fit (see below). For all preparations, this correction does not exceed 5\,\%. Furthermore, the effect of these losses on the measurements of $g_{nn}(r\neq0)$ is on the order of $p_{\sigma}\,\ll 1$. In practice, we consider occupation probabilities measured after blasting state $\ket{3}$ for which the blast procedure is more reliable (see above), with the exception of the point of largest $\eta$ in Fig.~\ref{fig:fig4}b, for which we use all blasted images to get better signal-to-noise ratio. For consistency the values of $k_{\rm{F}}$ are also corrected accordingly for equal-spin losses.

Finally losses due to $\uparrow\downarrow$-pairs can be calculated locally in small sub-regions of the cloud, allowing to probe $g_{ii, \uparrow\downarrow}$ for different interaction parameters in a single experimental image (see below).

\subsection*{Contact extraction}

\begin{figure}[t!]
    \centering
    \includegraphics[width=\columnwidth]{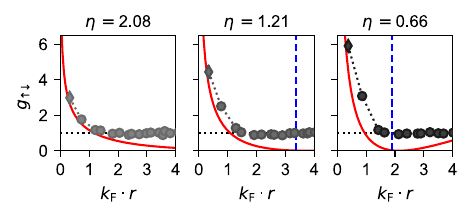}
	\caption{\textbf{Short range limit of $g_{\uparrow\downarrow}$}. Comparison of pair correlations (markers) and their short-range asymptotic behaviour Eq.~({\ref{eq:g2_C}}) (solid red line) where the contact density is obtained from the diamond. The vertical blue dashed line corresponds to $k_{\rm{F}}a$. On the BCS side and for our experimental parameters Eq.~({\ref{eq:g2_C}}) remains valid over the first few points of $g_{ij,\uparrow\downarrow}$. For stronger attraction the region of validity for  Eq.~({\ref{eq:g2_C}}) becomes smaller.}
    \label{fig:Short_range_g2}
\end{figure}

The values of the first points of $g_{ij,\uparrow\downarrow}$ give direct access to the contact density. At short-range, the functional form of $g_{\uparrow \downarrow}(r)$ is given in Eq.~(\ref{eq:g2_C}). On discretised space, this equation becomes: 
\beq
4\frac{C}{k_{\rm F}^4} {=} \frac{g_{ij,\uparrow\downarrow}}{\int_0^{+\infty}dr\ D_{ij}(r)\log ^2(\alpha r/a)}
\eeq 
To extract the contact we implement the above relation for $g_{ii,\uparrow\downarrow}$ (same-site correlator measured from losses) where the conditions of validity of Eq.~(\ref{eq:g2_C}), $r \ll a,\ k_{\rm{F}}^{-1}$, are best fulfilled. For data in the BCS regime where Eq.~(\ref{eq:g2_C}) is valid up to larger distances in real space, one can also extract the values of the contact density from the second and third points of $g_{ij,\uparrow\downarrow}$ which are consistent with one another. In Fig.~\ref{fig:Short_range_g2} we plot $g_{\uparrow\downarrow}$ and its short-range asymptotic behaviour where the contact density is determined from $g_{ii,\uparrow\downarrow}$. The errors shown in Fig.~\ref{fig:fig4}  take into account uncertainties on $g_{ij}$ and $a$.
\\

\subsection*{Effect of harmonic trapping potential}

In the $xy-$plane, the harmonic confinement leads to an inhomogeneous cloud. Within the LDA, small sub-regions of the cloud where density variations are small can be considered as homogeneous systems described by local thermodynamic variables (chemical potential $\mu(\textbf{r})$, interaction parameter $\eta(\textbf{r})$ and reduced temperature $T/T_{\rm{F}}(\textbf{r})$). Therefore probing different parts of the cloud is equivalent to probing many homogeneous 2D Fermi gases in the grand-canonical ensemble with varying interaction parameters and reduced temperatures.

At low temperatures, the contact density depends only weakly on $T/T_{\rm{F}}$ \cite{he2022}. Given our low temperatures, computing $C/k_{\rm{F}}^4$ over different parts of the cloud thus gives access to its behaviour over a wide range of interaction parameters while the variations of $T/T_{\rm{F}}$ only play a marginal role.

To get sufficiently high signal-to-noise ratio, we measure losses averaged over rings of finite thickness where density variations are typically on the order of a few percent. This is equivalent to measuring the contact density averaged over slightly different local interaction parameters. Given the curvature of $C/k_{\rm{F}}^4 = f(\eta)$, this leads to a small systematic upwards shift of the measured contact density when compared to the theoretical curve, which however remains smaller than our error bars. Additional sources of fluctuations on the interaction parameter, such as shot to shot fluctuations of the atom number, are also expected to contribute to this effect.

Similar considerations apply for all quantities calculated over quasi-homogeneous regions, such as two- and three-point density-density correlations. Errors on the interaction parameter include the uncertainty on $a$ as well as the variation of $k_{\rm{F}}$ over the region of interest and correspond to 95\,\% confidence intervals.

\subsection*{AFQMC simulations}

The AFQMC simulations presented in this work utilize 
state-of-the-art  finite-temperature determinantal quantum Monte Carlo methods, 
especially the speedup algorithm of Ref.~\cite{he2019}. 
This algorithm reduces the computational complexity from the conventional $\mathcal{O}(N_s^3)$ to $\mathcal{O}(N_s N_e^2)$, where $N_s$ and $N_e$ denote the number of lattice sites and fermions, respectively. This improvement is 
essential 
for interacting Fermi gases, where 
$N_s \gg N_e$ 
in order to approach the continuum limit. 
The large gain with the speedup algorithm enables us to reach sufficiently large $N_e$, in the continuum limit, for 
reliable comparison with experiment.

In AFQMC simulations, we model the uniform 2D Fermi gas with contact attraction by the attractive Hubbard model on a square lattice 
($N_s=L\times L$), 
with quadratic 
kinetic energy dispersion $\varepsilon_{\mathbf{k}}=k_x^2+k_y^2$ and a system-dependent on-site interaction $U$ determined from $\eta=\log(k_{\rm F}a)$~\cite{shi2015,he2022}. 

The two-point correlation functions $g_{nn}(r)$, $g_{\sigma\sigma}(r)$, and $g_{\uparrow\downarrow}(r)$ are measured in AFQMC with the standard procedure (see Refs.~\cite{he2019} and ~\cite{he2022}, including supplemental materials therein, for details.)

Since there are no equilateral triangles on a finite-size square lattice, we 
use an interpolation scheme to compute the three-point correlations $g_{\uparrow\uparrow\uparrow}$, $g_{\uparrow\uparrow\downarrow}$, and $g_{nnn}$ as depicted in Fig.\,\ref{fig:fig3}a of the main text. Specifically, we first measure the correlations on two isosceles triangles that share the same base, and then linearly interpolate between the two results to estimate the correlation on the corresponding equilateral triangle. For example, the numerical results on the two isosceles triangles $[(x=0,y=0)$-$(x=2,y=0)$-$(x=1,y=1)]$ and $[(x=0,y=0)$-$(x=2,y=0)$-$(x=1,y=2)]$ can be used to estimate the three-point correlations in the equilateral triangle of $[(x=0,y=0)$-$(x=2,y=0)$-$(x=1,y=\sqrt{3})]$. 

The AFQMC results for both the two-point and three-point correlations, presented in Fig.\,\ref{fig:fig2} and Fig.\,\ref{fig:fig3} of the main text, are obtained from a system with $L=45,N_e = 58$. These results are highly accurate, with relative errors on the order of $0.1\%$, and therefore the error bars are omitted from the plots for clarity. To ensure the reliability of these results, we have performed cross-checks with data from other system sizes, including $L = 25,N_e = 42$ and $L=35,N_e=74$, and have found excellent agreement among these results. Thus, the AFQMC data of the correlations in this work 
are effectively in 
the continuum and thermodynamic limits. The $T=0$ contact density results from AFQMC simulations presented in Fig.\,\ref{fig:fig4} of the main text are recovered from Fig.\,2 of Ref.~\cite{shi2015} and the Equation of State results therein.

\subsection*{Shape of equal-spin correlation functions}

\begin{figure}[!t]
    \centering
    \includegraphics[width=\columnwidth]{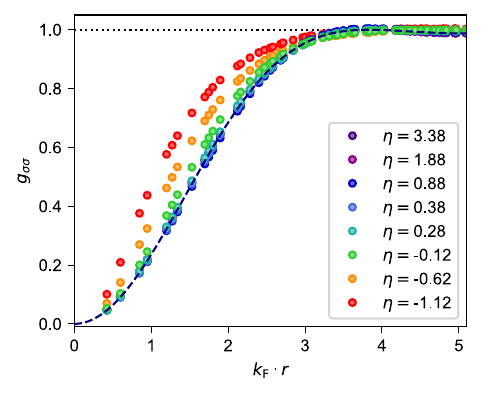}
	\caption{\textbf{Equal-spin correlations across the BEC-BCS crossover.} Equal-spin correlations obtained via AFQMC calculations at $T/T_{\rm{F}} = 0.125$. We also show the theoretical correlation function of an ideal Fermi gas at the same temperature (black dashed line). Simulated correlation functions remain very close to the prediction for an ideal Fermi gas above $\eta \gtrsim 0.2$.}
    \label{fig:g2_uu_AFQMC}
\end{figure}

\begin{figure*}[]
	\centering
	\includegraphics[width=0.8\textwidth]{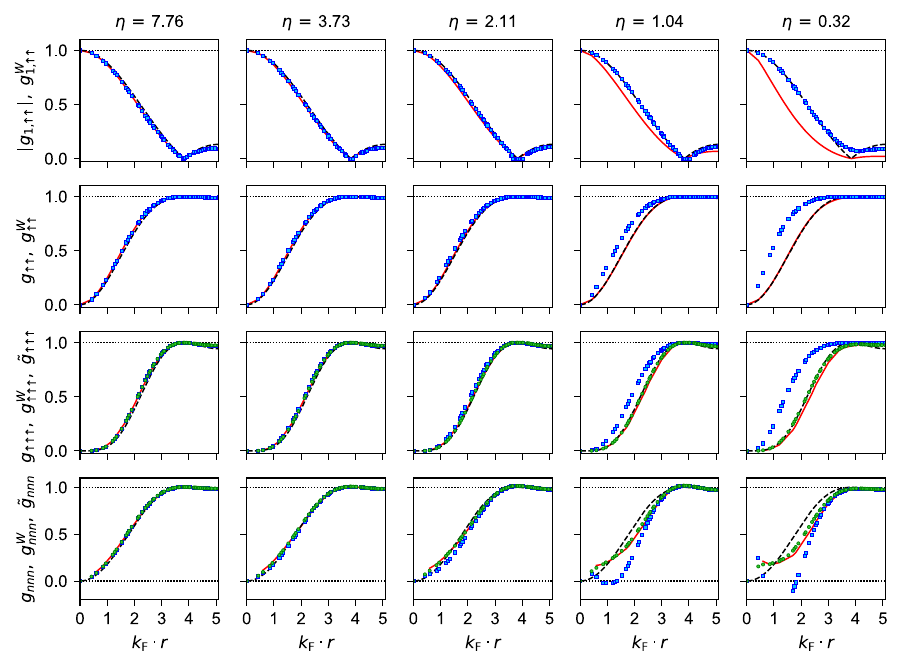}
	\caption{\textbf{Testing Wick relations in the 2D BEC-BCS crossover with AFQMC.} First row: $|g_{1,\up\up}|$ (red solid line) and $g^{\rm W}_{1,\up\up}=\sqrt{1-g_{\up\up}}$ (blue squares). Second row: $g_{\up\up}$ (red solid line) and $g^{\rm W}_{\up\up}=1-g_{1,\up\up}^2$ (blue squares). Third row:  $g_{\up\up\up}$ (red solid line), $g^{\rm W}_{\up\up\up}$ (blue squares), and $\tilde{g}_{\up\up\up}$ of Eq.~(\ref{eq:g3wick1}) (green circles). Fourth row: $g_{nnn}$ (red solid line), $g^{\rm W}_{nnn}$ (blue squares), and $\tilde{g}_{nnn}$ of Eq.~(\ref{eq:g3wick2}) (green circles). Wick relations involving $g_{1,\up\up}$ all fail as one moves away from the BCS side towards the crossover, but the relations Eqs.~(\ref{eq:g3wick1}--\ref{eq:g3wick3}) are satisfied to a high degree until the crossover. The black dashed curves are the $T=0$ prediction for a polarised Fermi gas (1$^{\rm st}$ to 3$^{\rm rd}$ row) and non-interacting spin-1/2 mixture (4$^{\rm th}$ row). The AFQMC are performed at $T/T_{F}=0.182$, $0.153$, $0.112$, $0.084$, $0.079$, $0.072$ from left to right, corresponding to the experimental parameters.}
	\label{fig:Wick_AFQMC}
\end{figure*}

One can check the functional form of $g_{\sigma\sigma}$ obtained from AFQMC calculations across the BEC-BCS crossover and compare them to the theoretical prediction for the ideal Fermi gas at the same temperature. Above $\eta\gtrsim0.2$, we find that equal-spin correlation functions are indistinguishable from the correlations in an ideal fully-polarised Fermi gas as shown in Fig.~\ref{fig:g2_uu_AFQMC}. This is in contrast to the prediction from BCS theory where the Fermi hole notably shrinks with increasing attraction strength.

\begin{figure*}[h]
    \centering
    \includegraphics[width=\textwidth]{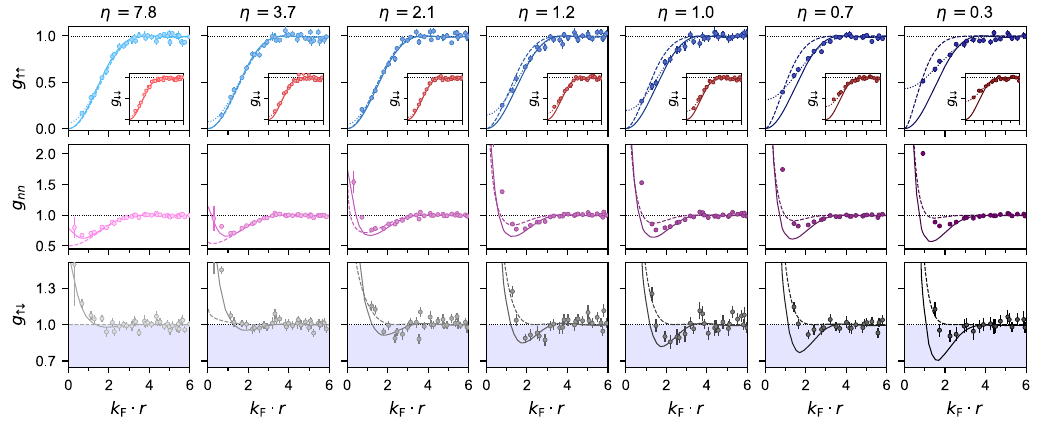}
    \caption{\textbf{Two-point charge- and spin-  correlations across the BEC-BCS crossover}. (Top row): equal-spin correlations. (Middle row): Charge correlations. (Bottom row): Pair correlations. The thin diamonds are obtained by measuring atom losses. The dashed line is the prediction of mean-field BCS theory. Solid lines are AFQMC calculations performed for our experimental parameters. The dotted lines on the top row correspond to a two-parameter fit of $g_{\sigma\sigma}$.}
    \label{fig:figS6}
\end{figure*}

This motivates the functional form of the two-parameter fit we use for our experimental  $g_{ij,\up\up}$. It is based on the correlation function of a quasi-2D ideal Fermi gas at finite temperature with several populated transverse motional states. In an ideal Fermi gas, transverse levels can be populated via two mechanisms: temperature and Fermi energy (see \cite{jongh2024}). A key difference here is that interactions can also induce non-zero populations in the excited $z$-levels. Therefore a model where temperature is left as the sole free fitting parameter and determines populations in excited states via the Fermi-Dirac distribution (as is done for the non-interacting case) fails to reproduce the data for strong interactions. Instead, we find that a two-parameter fit where both $T$ and the population in excited levels are left as independant quantities allows to closely match the experimental data (see Figs.~\ref{fig:fig2} and \ref{fig:figS6}). For simplicity and to avoid possible overfitting, we only allow for two different transverse levels to be populated (with populations $p$ and $1-p$). The explicit formula reads:

\beqa
g_{\sigma\sigma}^{\rm tot}(r) &=& p^2g_{\up\up} \left( \sqrt{p}k_{\rm{F}}r, \frac{T}{pT_{\rm{F}}} \right) \nonumber\\
&+& (1-p)^2g_{\up\up} \left(\sqrt{1-p}k_{\rm{F}}r, \frac{T}{(1-p)T_{\rm{F}}}\right)\nonumber\\
&+& 2p(1-p).
\eeqa

To obtain the best signal-to-noise ratio, we use the result of this fit to calculate $g_{\up\down}$. We also use it to estimate $g_{ii,\sigma\sigma}$, and thus same-spin losses (see above). Note that this functional form is purely phenomenological, and that the value of $T$ obtained from the fit does not correspond to any meaningful temperature.

\subsection*{Wick Analysis}

For fermionic systems described by a quadratic hamiltonian, Wick's theorem relates the coherence function $g_{1}(r)$ to higher-order correlations. In the following, all relations written for three-point correlations assume equilateral triangles. In an ideal fully polarised Fermi gas, one has:
\beqa
g^{\rm W}_{\up\up}(r) &=& 1 - g_{1,\up\up}^2(r)\label{eq:Wick_order2}\\
g^{\rm W}_{\up\up\up}(r) &=& 1 - 3g_{1,\up\up}^2(r) + 2g_{1,\up\up}^3(r),
\label{eq:Wick_order3}
\eeqa where the index 1 denotes the first order correlation function $g_{1,\up\up}(r)=\langle\Psi_\up^\dagger(r)\Psi_\up(0)\rangle/n$. 
For an interacting spin-$1/2$ mixture within the framework of mean-field BCS theory, Wick theorem also applies. The two previous equations still hold as well as the following ones:
\beqa
g^{\rm W}_{nnn}(r) = &\frac{3}{2}&\Big[g_{\up\up}(r)\Big (1+g_{1,\up\up}(r)\Big ) \nonumber\\
&+&\, g_{\up\down}(r)\Big (1-g_{1,\up\up}(r)\Big )\Big] \nonumber\\
&+&\, 2g_{1,\up\up}^{3}(r) - 2 \label{eq:Wick_nnn}\\
g^{\rm W}_{\up\up\down} &=& g_{\up\up}(r)+2g_{\up\down}(r) \nonumber \\
&+& 2\left[ g_{\up\up}(r) - g_{\up\down}(r) \right]g_{1,\up\up}(r) \nonumber \\
&+& 2g_{1,\up\up}(r)^{3}-2.
 \label{eq:Wick_uud}
\eeqa

The three previous relations and Eqs.~(\ref{eq:g3wick1}-\ref{eq:g3wick3}) of the main text are equivalent assuming that Wick's theorem at first order, Eq.~(\ref{eq:Wick_order2}) is verified.

Experimental images of the spin- and charge-densities do not give access to the coherence function of the sample. Instead, we compute \mbox{$g^{W}_{1,\up\up}(k_{\rm{F}}r) = \pm\sqrt{1\,-\,g_{\up\up}(k_{\rm{F}}r)}$} (the sign is positive at distances below the size of the Fermi hole $k_{\rm F}r\lesssim 3.8$ which are the most relevant experimentally). We compare the measured values of $g_{\up\up\up}$, $g_{nnn}$ and $g_{\up\up\down}$ over equilateral triangles to the Wick-inspired prediction of Eqs.~(\ref{eq:g3wick1}-\ref{eq:g3wick3}) which only involves quantities that can be directly measured from experimental images. This is shown in Fig.~(\ref{fig:fig3}) of the main text.

In AFQMC simulations, Wick's prediction can be directly tested since all correlations can be computed (with growing cost as the order of the correlation increases). We show in Fig.~\ref{fig:Wick_AFQMC} that the Wick relation between $g_{\up\up}$ and $g_{1,\up\up}$ breaks down in the strongly-interacting regime. However computing $g^W_{1,\up\up}$ and using this quantity instead of the true $g_{1,\up\up}$ to compute correlations of higher order (as is done with the experimental data) yields a remarkably close match between the data and Eqs.~(\ref{eq:g3wick1}--\ref{eq:g3wick3}). It is interesting to note that while increasing interactions lead to a broadening of the Fermi surface, indicated by a narrower peak of $g_{1,\up\up}$, the antibunching of fermions encoded in the Fermi hole in $g_{\up\up}$ remains robust up to $\eta\sim0$.

\subsection*{Mean-Field BCS correlations}
Within mean-field BCS theory, the functional form of correlations can be derived from the quasi-particle amplitudes and dispersion relation in combination with Wick's theorem, which is valid in this framework. The final result comes out to \cite{obeso-jureidini2022}:
\beqa
g_{\uparrow\uparrow}(r) &=& 1\, -\, \left\lvert \frac{2}{k_{\rm{F}}a} J_1(k_{\rm{F}}r) K_1(r/a) \right\rvert ^2\\
\label{eq:MF_g_uu}
g_{\uparrow\downarrow}(r) &=& 1\, +\, \left\lvert \frac{2}{k_{\rm{F}}a} J_0(k_{\rm{F}}r) K_0(r/a) \right\rvert ^2
\label{eq:MF_g_ud}
\eeqa
Third-order correlations are computed likewise using Wick's theorem. Eq.~(\ref{eq:MF_g_ud}) satisfies the correct short-range behaviour of $g_{\uparrow\downarrow}(r)$ with $C/k_{\rm{F}}^4 = 1/(k_{\rm{F}}a)^2$.

\subsection*{Fermi liquid theory}

For weakly attractive 2D Fermi gases, the equation of state of the gas can be expanded in terms of $\eta$ as \mbox{$E/N = E_{FG}[1-\frac{1}{\eta}+\frac{A}{\eta^2}+O(\frac{1}{\eta^3})]$} where \mbox{$A = 3/4-\ln{(2)}$} \cite{engelbrecht1992}. The equation of state is related to the contact density by \cite{werner2012}:
\beq
\frac{C}{k_{\rm{F}}^4} = \frac{1}{4E_{FG}}\times\frac{d (E/N)}{d\eta}
\eeq

Therefore, 
\beq
\frac{C}{k_{\rm{F}}^4}  = \frac{1}{4}\times\left(\frac{1}{\eta^2}-\frac{2A}{\eta^3}\right).
\eeq

\subsection*{Extended Data}
Below, we show an extended version of Fig.~\ref{fig:fig2}b and Fig.~\ref{fig:fig3}b.

\begin{figure*}[b]
    \centering
    \includegraphics[width=\textwidth]{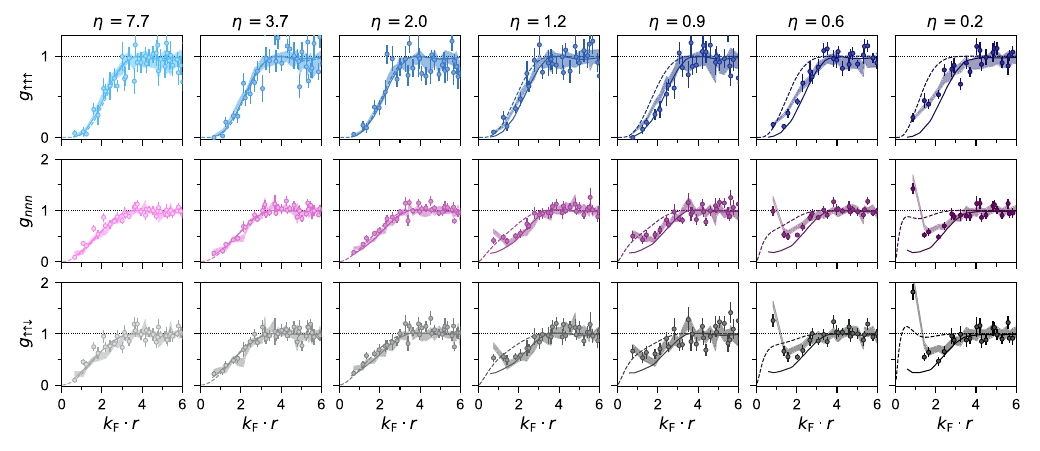}
    \caption{\textbf{Three-point charge- and spin-correlation functions}. (Top row): equal-spin correlations. (Middle row): Charge correlations. (Bottom row): Inter-spin correlations. The shaded areas correspond to the prediction of Eqs.~(\ref{eq:g3wick1}--\ref{eq:g3wick3}). We show the prediction of mean-field BCS theory (dashed lines), and the results of AFQMC calculations performed for our experimental parameters (solid lines).}
    \label{fig:figS7}
\end{figure*}

\end{document}